# Resonant Coherent Phonon Generation in Single-Walled Carbon Nanotubes through Near-Band-Edge Excitation

Yong-Sik Lim[1*], Jae-Geum Ahn[1], Ji-Hee Kim[2], Ki-Ju Yee[2], Taiha Joo[3], Sung-Hoon Baik[4], Erik H. Hároz[5], Layla G. Booshehri[5], and Junichiro Kono[5*]

[1]Department of Applied Physics, Konkuk University, Chungju, Chungbuk 380-701, Republic of Korea

[2]Department of Physics, Chungnam National University, Daejon 305-764, Republic of Korea

[3]Department of Chemisty, POSTECH, Pohang 790-784, Republic of Korea

[4]Quantum Optics Research Division, Korea Atomic Energy Research Institute, Daejon 305-353, Republic of Korea

[5]Department of Electrical and Computer Engineering, Rice University, Houston, Texas 77005, USA

[*]Corresponding authors

**Abstract**  We have observed large-amplitude coherent phonon oscillations of radial breathing modes (RBMs) in single-walled carbon nanotubes excited through the lowest-energy ($E_{11}$) interband transitions.  In contrast to the previously-studied coherent phonons excited through higher-energy ($E_{22}$) transitions, these RBMs show comparable intensities between ($n - m$) mod 3 = 1 and −1 nanotubes.  We also find novel non-resonantly excited RBMs over an excitation range of ~300 meV above the $E_{11}$ transition, which we attribute to multi-phonon replicas arising from strong exciton-phonon coupling.

Single-walled carbon nanotubes (SWNTs) are one of the most ideal one-dimensional systems available today for studying the effects of dimensionality on confined carriers and phonons and their mutual interactions.[1,2]  Advances in optical studies such as photoluminescence excitation (PLE) spectroscopy have led to definitive assignments of spectral features to specific chiralities, or ($n,m$).[3-5]  Recent theoretical and experimental studies illuminate the importance of pronounced excitonic effects in interband optical processes in SWNTs due to their one-dimensionality.[6-10]  Furthermore, PLE microscopy, polarized PLE, and photoconductivity studies have revealed a variety of phonon-assisted peaks,[11-13] suggesting strong exciton-phonon coupling.

Recently, we have reported the observation of coherent phonon (CP) oscillations of radial breathing modes (RBMs) in SWNTs induced by impulsive $E_{22}$ optical transitions.[14-16]  We found that ($n - m$) mod 3 ≡ $\nu = -1$ tubes have much larger intensities than $\nu = +1$ tubes, the same trend as seen in resonant Raman spectroscopy studies.[17-19]  We also showed that CP spectroscopy has several advantages over Raman spectroscopy, including no Rayleigh scattering and PL backgrounds.  Here, we use these advantages to study CP oscillations of RBMs in smaller-diameter SWNTs, which showed $E_{11}$ transitions within the wavelength range accessible with a Ti:Sapphire laser. The data displayed several RBMs resonantly excited by the $E_{11}$ and $E_{22}$ transitions, respectively, as well as non-resonantly excited novel RBMs found over widespread intermediate excitations between the $E_{11}$ and $E_{22}$ optical transitions. A linear square fitting analysis method known as the linear prediction based on singular value decomposition (LPSVD),[20,21] which retrieves time constants shorter than the duration of the excitation pulse in the presence of noise, was applied for obtaining the resonance excitation profile to compare the magnitude of vibration strengths of resonantly-excited RBMs between $v= +1$ and $v= -1$ semiconducting nanotubes.

The sample used was an aqueous solution containing micelle-suspended CoMoCAT SWNTs in 1% dodecyl benzene sulfate (SDBS) in $D_2O$ inside a 1-mm quartz cell. Multiple RBMs of SWNTs corresponding to different diameters were simultaneously excited within the broad bandwidth of femtosecond pulses from a Ti:Sapphire oscillator with a repetition rate of 90 MHz and an average power of 300-400 mW through degenerate pump-probe, differential absorption spectroscopy. The pump and probe beams were divided by a beam splitter with a 7:3 ratio and was focused by a lens with a 5-cm focal length. A portion of the probe beam acted as the reference beam for one photodiode of a Nirvana balance detector, while the transmitted probe was aligned to the other photodiode. We tuned the center wavelength of the pump beam over a wide wavelength range of 720–1000 nm in steps of 5 nm to investigate the $E_{11}$ and $E_{22}$ transitions.[22] To get a high signal-to-noise ratio, we averaged multiple signals with a fast scanner of 20 Hz and a high speed data acquisition card. All signals were averaged over 10,000 scans at room temperature.

Figures 1(a) and 1(b) show CP oscillations of RBMs resonantly excited through $E_{11}$ and $E_{22}$ optical transitions at selected excitation wavelengths within the 720-1000 nm range. Each trace shows a strong beating pattern due to the simultaneous excitations of multiple RBMs, which sensitively changes with the photon energy, implying that the CP oscillations are dominated by a few resonantly-excited RBMs. It should be noted that, at long wavelengths [Fig. 1(b)], the data show a normalized differential transmission ($\Delta T/T$) of the order of ~$10^{-4}$ near time zero, which is 2-3 times larger than that for short-wavelength excitation [Fig. 1(a)], implying that the CoMoCAT sample consists of nanotubes with dominantly smaller diameters.

Figure 2(a) shows contour plots of the CP intensity in log scale as a function of excitation photon energy between 1.23 eV (1000 nm) and 1.71 eV (720 nm) and RBM frequency between 155 cm$^{-1}$ and 400 cm$^{-1}$, obtained through a LPSVD analysis (see Online Support Information for more details). Here, rectangular symbols represent $E_{22}$ transitions and triangular symbols represent $E_{11}$ transitions.[3] Red and white symbols are for $\nu = -1$ tubes, while blue and black symbols are for $\nu = 1$ tubes. Figures 2(b) and 2(c) show representative CP spectra, corresponding to horizontal cuts of the contour map in 2(a). Figure 2(b) shows CP spectra for excitation wavelengths of 720-740 nm with a step size of 5 nm, exhibiting three dominant RBMs between 240 cm$^{-1}$ and 280 cm$^{-1}$, all of which are $E_{22}$-excited, $\nu = -1$ tubes. Specifically, they belong to the $2n + m = 22$ family – i.e., (11,0)/(10,2), (9,4), and (8,6) tubes having frequencies of 267 cm$^{-1}$/266.1 cm$^{-1}$, 258.3 cm$^{-1}$, and 246.6 cm$^{-1}$, respectively. The peak at 306 cm$^{-1}$ is primarily due to the $E_{22}$-excited RBM of (9,1) tubes ($E_{22}$ ~ 1.76 eV), with some contribution from (6,5) tubes excited between $E_{11}$ and $E_{22}$ (as described later). The black curve in Fig. 2(c) was taken with 765 nm (1.62 eV) excitation, showing $E_{22}$-excited CPs for the $2n + m = 25$ family [(12,1), (11,3), and (10,5)] with frequencies between 220 cm$^{-1}$ and 240 cm$^{-1}$, as well as for the $2n + m = 28$ family [(14,0)/(13,2) and (12,4)] at frequencies between 200 cm$^{-1}$ and 220 cm$^{-1}$.

The red trace in Fig. 2(c) is a CP spectrum measured at an excitation wavelength of 965 nm (1.28 eV), which shows two sharp RBMs at 307.5 cm$^{-1}$ and 329.7 cm$^{-1}$ corresponding to (6,5) and (7,3) tubes, respectively. These modes are excited through the $E_{11}$ optical transition, and their linewidths are 1.5 cm$^{-1}$ and 1.8 cm$^{-1}$, respectively. The linewidth of the (9,1) tube at 304.5 cm$^{-1}$ is as small as that of the (6,5) tube, whereas the RBM of the (8,3) tube at 297.6 cm$^{-1}$ shows a large linewidth of 3.8 cm$^{-1}$, similar to the (9,4) tube excited through the $E_{22}$ optical transition. Thus, apparently, the spectral linewidths of the RBMs are independent of the electronic resonances involved in the generation of the CP. In a third-order nonlinear optical experiment such as transient absorption spectroscopy employed here, phonon modes both in the ground and excited states ($E_{11}$ or $E_{22}$) contribute to the signal. Since the lifetime of the $E_{22}$ (and $E_{11}$) state is much shorter than the dephasing times of the RBMs, a transient absorption signal for times longer than ~1 ps consists of the ground state contribution exclusively.[23] That is, the phonon modes in the ground state are recorded almost exclusively in the present experiment, and therefore, the RBM linewidths should not be dependent on the electronic resonances involved.

To deduce quantitative information on how the CP signal changes with ν, chiral angle, diameter, and optical transitions ($E_{22}$ vs. $E_{11}$), we fully analyzed the resonance excitation profiles of the observed features in Fig. 2(a), taking into account the double-peak line-shape arising from the first derivative of a Lorentzian (see Ref. [14] as well as Online Supporting Information for more details). Table 1 summarizes the results obtained from such analysis, showing the CP strength, the resonance energy, the FWHM, the chirality, and the ν of each RBM feature. There are several distinguished characteristics and trends in Table 1 that are worth discussing. First, we find contrasting results between $E_{22}$-excited and $E_{11}$-excited CPs: for $E_{22}$-excited CPs (i.e., those excited at photon energies higher than 1.45 eV), $\nu = -1$ nanotubes show markedly higher intensities than $\nu = 1$ nanotubes (the latter are nearly invisible in the spectra), whereas for $E_{11}$-excited CPs (i.e., those excited at photon energies lower than 1.45 eV), the intensities are comparable between the $\nu = -1$ and $\nu = 1$ nanotubes. The $E_{11}$-excited RBMs for $\nu = 1$ tubes [(7,3), (6,5), and (5,4) tubes] are very strong. The RBM peak due to the (7,3) tube (diameter = 0.7 nm) is much stronger than the (8,3) tube ($\nu = -1$ tube with a comparable diameter of 0.78 nm). Second, we note that, for $E_{22}$-excited tubes, the CP intensity varies strongly within each family as a function of chiral angle. Namely, the intensity tends to decrease as the chiral angle increases (going from zigzag to armchair). This trend is the same as we observed earlier for HiPco samples[14] and can be explained through the chiral-angle dependence of the coupling matrix element between excitons and phonons.[24-26] Third, we see that the intensity of $E_{22}$-excited CPs decreases as we go from a low family index to a high family index, which is opposite to what we observed for HiPco tubes in the same Ti:Saphire spectral range.[14] This is due to the difference in the diameter distribution: our CoMoCAT SWNTs have an average diameter of 0.7 nm, while the average diameter of our HiPco SWNTs was 1.0 nm.[14] Fourth, the intensity of $E_{11}$-excited RBMs appears to have the same type of chiral-angle dependence as $E_{22}$-excited RBMs within the same family. This can be most clearly seen by comparing (6,5) and (7,3) tubes (both are family 17 tubes). Although (6,5) is the most populous species in our ensemble sample, its intensity is smaller than that for (7,3) tubes. Fifth, $\nu = -1$ nanotubes such as (9,1) and (8,3) tubes show CP intensities that are independent of whether they are through the $E_{11}$ or $E_{22}$ optical transitions. These results are partially consistent with the previous theoretical predictions [24-26] and close to our recent microscopic CP theory,[16] but they serve to invite more accurate theoretical calculations.

Finally, we discuss our surprising observation of non-resonantly excited strong RBMs of (6,5) and (7,3) tubes over a wide excitation range between the $E_{11}$ and $E_{22}$ transitions, indicated by the white ellipse in Fig. 2(a). Figure 3(a) shows CPs excited at 800 nm (1.54 eV). Compared with our previous CP study on HiPco samples, we observe several vibration modes from CoMoCAT samples at frequencies higher than 280 cm$^{-1}$. We can readily assign the mode at 373 cm$^{-1}$ to the RBM of the (5,4) tube with a diameter of 0.62 nm excited through the $E_{11}$ optical transition. In addition, we observe non-resonantly excited RBMs at 304 cm$^{-1}$, 307 cm$^{-1}$, and 330 cm$^{-1}$, corresponding to (6,5), (9,1), and (7,3) tubes, respectively. Furthermore, these spectral features are large and spread enough to surpass resonantly-excited RBMs over the entire excitation range of more than 300 meV between the $E_{11}$ and $E_{22}$ transition energies of (6,5) and (7,3) tubes. These non-resonantly-excited RBMs are also observed weakly for the (8,3) tube and likely to exist on the higher energy side of the $E_{22}$ optical transition for some of the tubes of the $2n + m = 22$ and $2n + m = 25$ families, as shown in Fig. 2(a).

To investigate this spectral range more closely, we performed a PLE study on our sample over the excitation range of 775-945 nm (1.59-1.30 eV), as shown in Fig. 3(b). In the emission wavelength range of 885.5-1215.7 nm (1.39-1.01 eV), strong PL from the (6,5) tube, as well as weak PL from (7,5), (8,3), and (8,4) tubes, is observed. Different from an earlier report,[27] we do not find PL associated with two-phonon processes involving LO/LA phonons near the K edge, but instead find the contribution associated with two-phonon processes involving TA/TO phonons near the K or M edge with the assistance of a resonant Raman effect forming a straight dotted line across the excitation and emission energy range. It should be noticed in Figs. 3(b) and 2(a) that the peak position of the dominant PLE signal for the (6,5) tube near 1.45 eV coincides with the discontinuous position in a resonance excitation profile (vertical line) of RBMs for (9,1)/(6,5) tubes, which also holds true for the (7,3) tube. Therefore, we believe that the non-resonantly-excited strong CP signals between the $E_{11}$ and $E_{22}$ transitions can be attributed to many phonon-assisted emission processes above the $E_{11}$ transition, which explains the observation of non-resonant RBMs appearing over almost the entire excitation range between the $E_{11}$ and $E_{22}$ transitions.

In conclusion, we have studied coherent phonon oscillations of radial breathing modes in micelle-suspended CoMoCAT single-walled carbon nanotubes with small diameters in the near infrared from 1.23 to 1.71 eV. Unlike HiPco SWNTs, CoMoCAT SWNTs showed strong RBM CPs of semiconducting tubes resonantly excited through both $E_{11}$ and $E_{22}$ transitions within our excitation range. For excitation energies higher than 1.45 eV, semiconducting tubes of $\nu = -1$ dominantly showed resonant RBMs excited through the $E_{22}$ transition, showing similar results to our previous study on HiPco samples. For excitation energies lower than 1.45 eV, we observed several RBMs of semiconducting tubes of both $\nu = -1$ and $\nu = +1$ types. We calculated relative spectral intensities by analyzing the resonance excitation profiles to demonstrate that the CP intensity shows the same chiral-angle dependence in both types. We also found that the CP intensity of $\nu = +1$ tubes varied strongly depending on whether it was excited through $E_{11}$ or $E_{22}$ transition, while $\nu = -1$ tubes did not show such variation. Finally, we observed novel non-resonantly-excited strong RBMs over a wide excitation range of more than 300 meV between the $E_{11}$ and $E_{22}$ transitions of (6,5) and (7,3) tubes, which we identified as multiple phonon-assisted excitonic transitions above the $E_{11}$ transition and confirmed them through PLE measurements.

This work was supported by the Korea Science and Engineering Foundation (KOSEF) grant funded by the Korean Government (Most) (R01-2007-000-20651-0, 2008-03535, 2009-0085432), DOE-BES (through Grant No. DEFG02-06ER46308), the Robert A. Welch Foundation (Grant No. C-1509), and the National Science Foundation (OISE-0530220).

| (n,m) | 2n+m | $\nu$ | $E_{ii}$ | $\omega_{RBM}$ (cm$^{-1}$) | $\Gamma$ (meV) | $E_0$ (eV) | $I_0$ (a.u.) |
|---|---|---|---|---|---|---|---|
| (11,0) | 22 | -1 | $E_{22}$ | 267 | 81.7 | 1.714 | 3.27 |
| (10,2) | 22 | -1 | $E_{22}$ | 266.1 | 77.5 | 1.713 | 2.83 |
| (9,4) | 22 | -1 | $E_{22}$ | 258.3 | 66.7 | 1.728 | 2.34 |
| (8,6) | 22 | -1 | $E_{22}$ | 246.6 | 73.6 | 1.732 | 1.35 |
| (12,1) | 25 | -1 | $E_{22}$ | 238.5 | 50.2 | 1.555 | 0.95 |
| (11,3) | 25 | -1 | $E_{22}$ | 231.9 | 51.2 | 1.565 | 0.43 |
| (10,5) | 25 | -1 | $E_{22}$ | 227.1 | 42.1 | 1.564 | 0.32 |
| (14,0) | 28 | -1 | $E_{22}$ | 215.1 | 63.9 | 1.434 | 0.33 |
| (13,2) | 28 | -1 | $E_{22}$ | 210.9 | 54.5 | 1.432 | 0.24 |
| (12,4) | 28 | -1 | $E_{22}$ | 208.8 | 54.0 | 1.432 | 0.16 |
| (9,1) | 19 | -1 | $E_{11}$ | 304.5 | 27.2 | 1.352 | 5.28 |
| (8,3) | 19 | -1 | $E_{11}$ | 297.6 | 52.4 | 1.286 | 2.55 |
| (7,3) | 17 | 1 | $E_{11}$ | 329.7 | 54.2 | 1.248 | 4.34 |
| (6,5) | 17 | 1 | $E_{11}$ | 307.5 | 27.4 | 1.281 | 3.37 |
| (5,4) | 14 | 1 | $E_{11}$ | 372.9 | 34.3 | 1.49 | 0.64 |

**FIGURE AND TABLE CAPTIONS**

**Figure 1** Coherent phonon oscillations measured at center wavelengths of (a) 720, 760, 800, 830, and 860 nm and (b) 880, 910, 940, 970, and 1000 nm, using degenerate pump and probe pulses with a bandwidth of 30-40 nm and a pulse-width of 40-50 fs.

**Figure 2** (a) 2-D log plot of a Fourier transform of CP oscillations, retrieved through the LPSVD method, measured over a photon energy range of 720-1000 nm (1.71-1.23 eV) with a 5-nm step size. The rectangular symbols stand for $E_{22}$ transitions and triangle symbols for $E_{11}$ transitions. Both yellow and black symbols are for $v = +1$, while red and blue are for $v = -1$. (b) CP spectra measured at center wavelengths of 720, 725, 730, 735, and 740 nm, showing RBMs resonantly excited through $E_{22}$ optical transitions. (b) Comparison of RBMs resonantly excited through $E_{22}$ and $E_{11}$ optical transitions with degenerate pump-probe pulses with center wavelengths at 765 nm and 965 nm, respectively.

**Figure 3** (a) CP spectrum measured with a center wavelength of 800 nm (1.54 eV). Non-resonantly-excited strong RBMs are indicated by blue letters. (b) PLE spectra for (6,5) tubes in the same CoMoCAT sample as used in the CP experiments. The dotted linear lines indicate emissions via resonant Raman processes with multiple phonon bands such as G-band, M-band, and D-band.

**Table 1** Summary of observed coherent phonons of the radial breathing mode. $(n,m)$: chirality indices, $2n + m$: chirality family, $v = (n - m)$ mod 3: chirality type, $E_{ii}$: transition type, $\omega_{RBM}$ (cm$^{-1}$): phonon frequency, $\Gamma$ (meV): full width at half maximum (FWHM) of excitation profile, $E_0$ (eV): transition energy, and $I_0$: calculated Lorentzian strength.

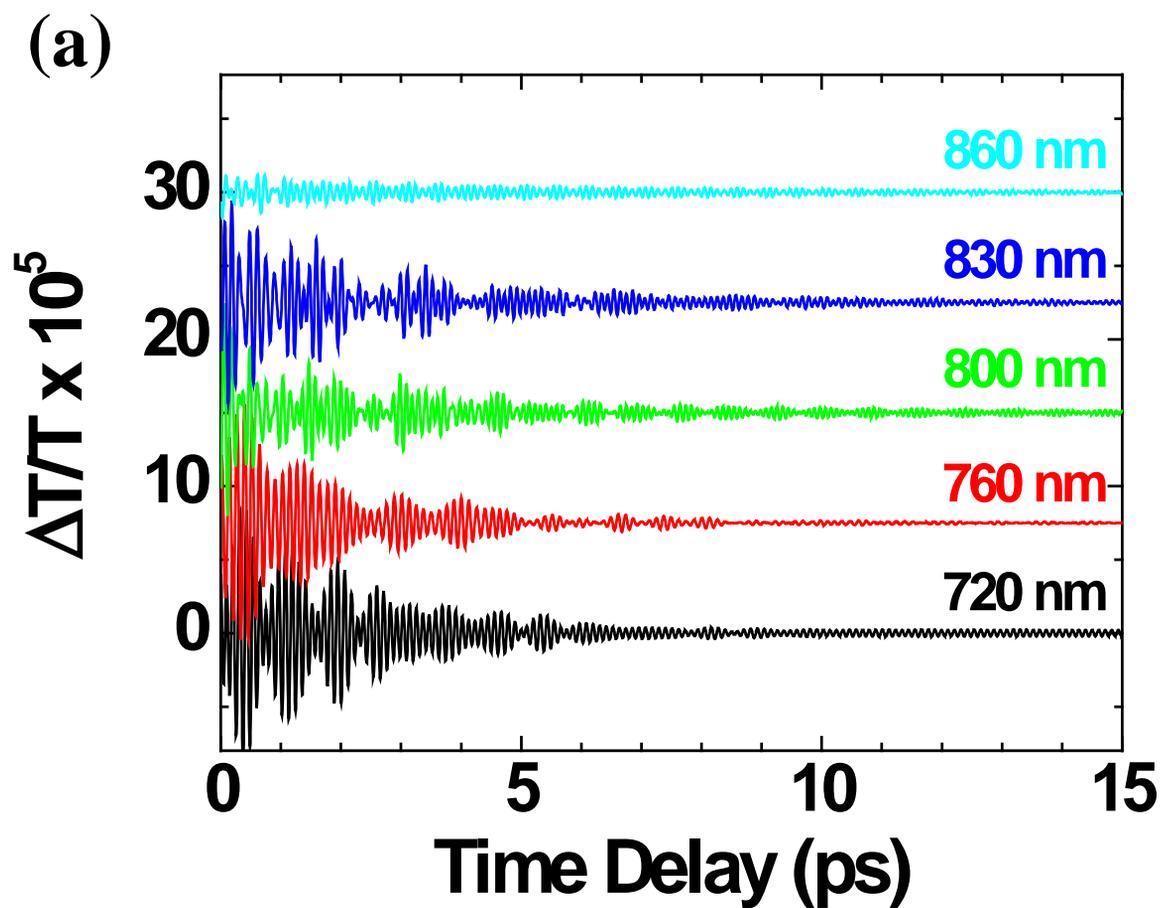
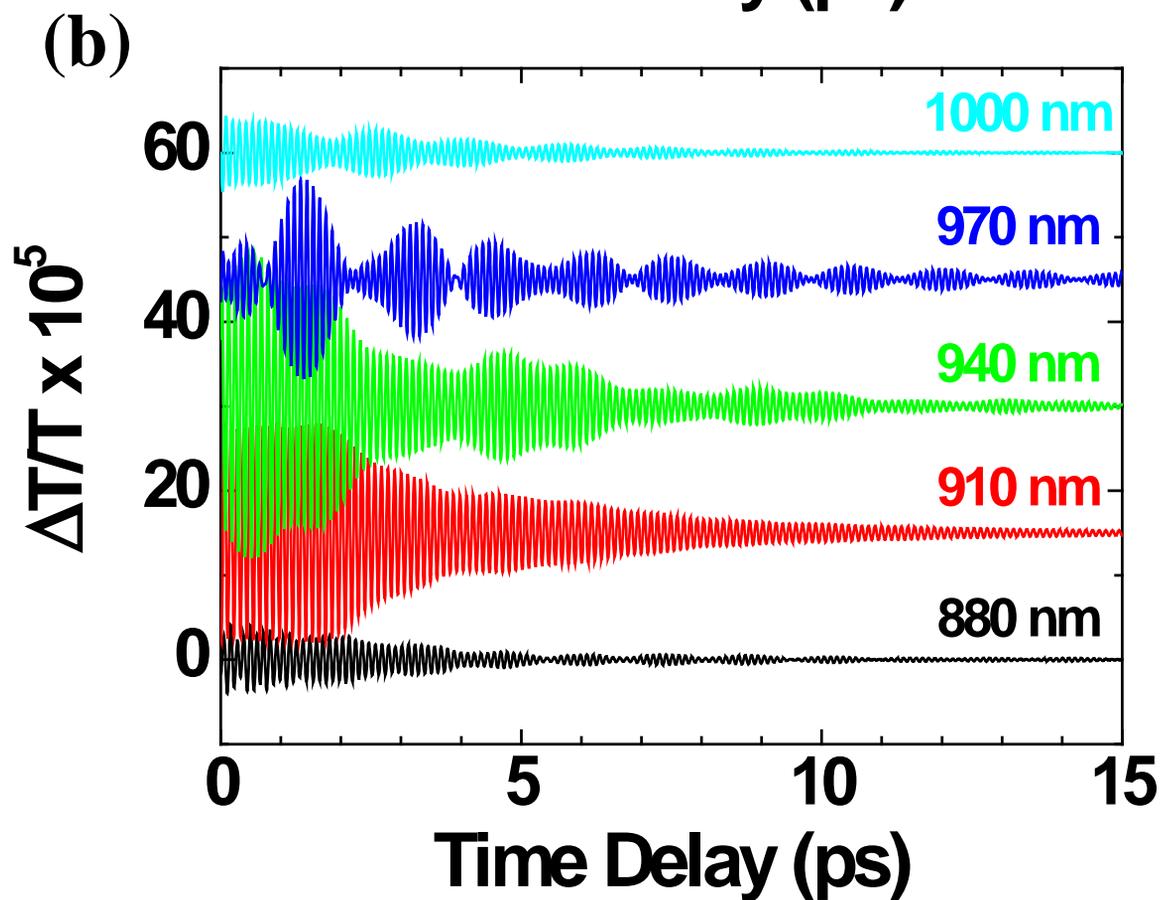

Fig. 1

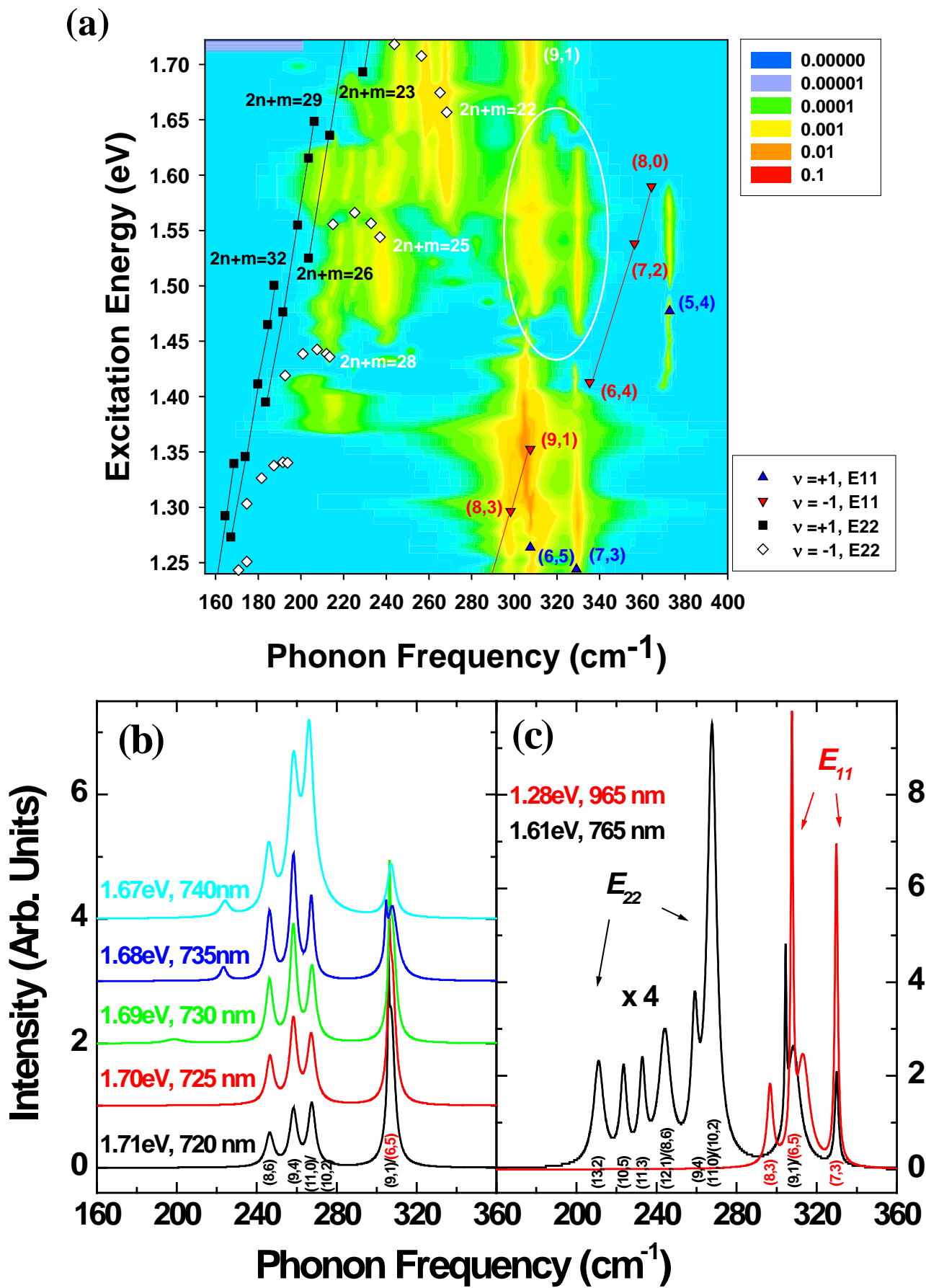

Fig. 2

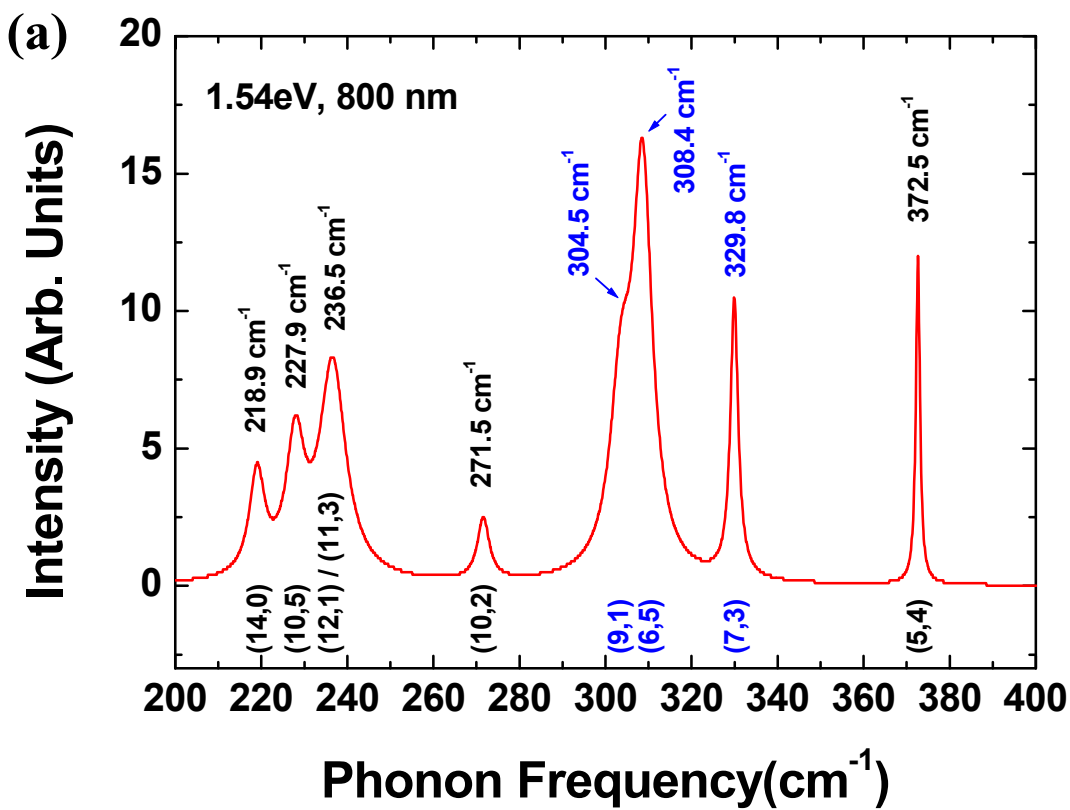

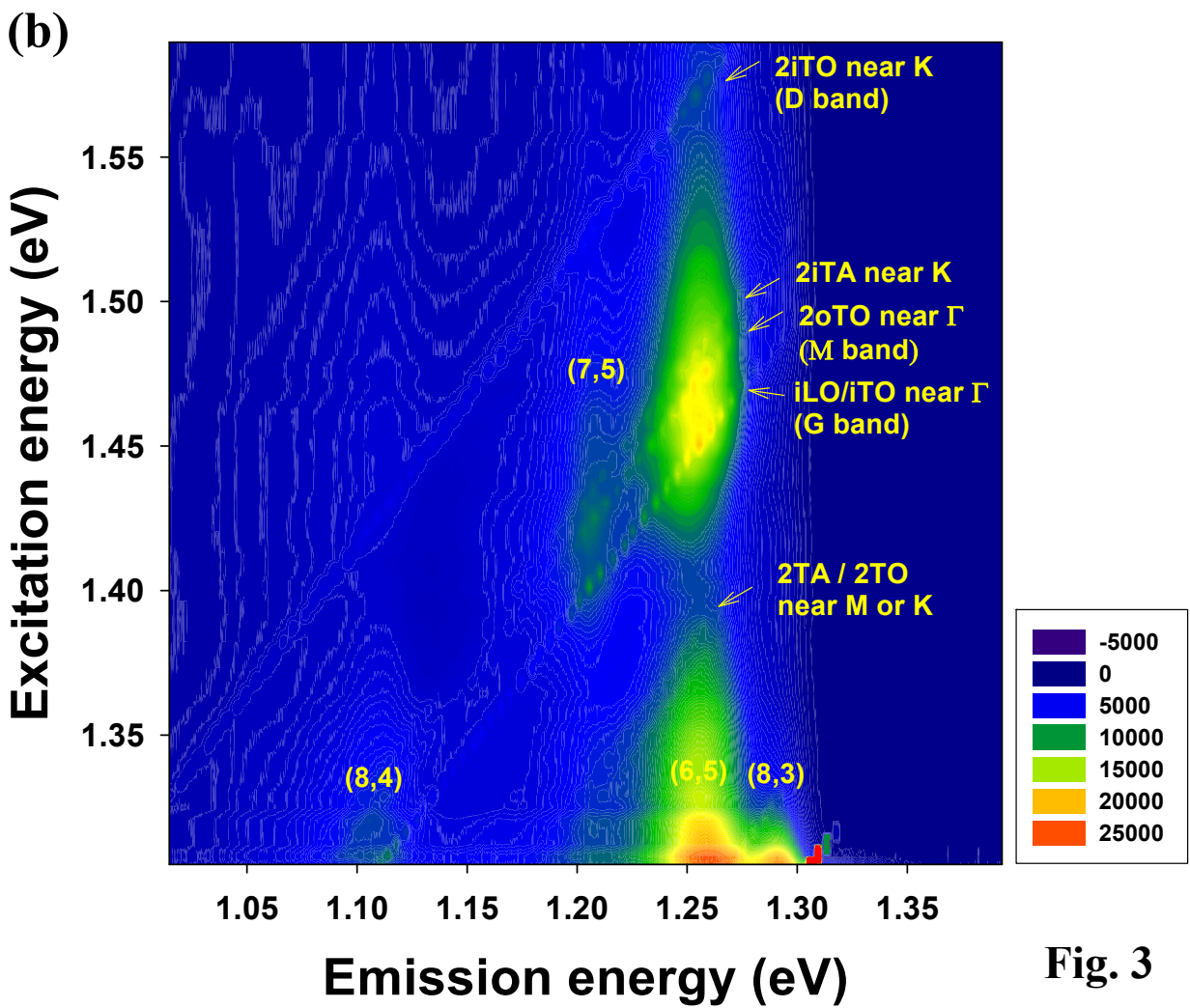

Fig. 3

# Supporting Information

Here we provide more information on the Linear Prediction based on Singular Value Decomposition (LPSVD) Method that we used to retrieve time constants shorter than the duration of the excitation pulse and obtain resonance excitation profiles for coherent phonons.

The LPSVD is an alternative method to fast Fourier transformation (FFT) for analyzing time-domain signals [S1-S3]. It has been demonstrated to give better results than FFT in many cases [S4-S5]. In LPSVD, a time domain signal is linearized by assuming that the signal is a sum of exponentials and damped sinusoids. Amplitudes, frequencies, time constants, and phase are obtained directly from linear algebra, and the results are exact within the assumption. For components with their amplitudes comparable to the noise of the signal, the result of a LPSVD-based analysis is model dependent, and care must be exercised in determining the number of components in the signal.

The strength ($I_0$) of each CP feature was calculated by fitting its excitation profile with the first derivative of a Lorentzian function [S-6]. Figure S-1 shows some fitting examples for a few chiral nanotubes with resonance excitation profiles obtained by LPSVD and FFT. There is no notable difference in fitted results between the LPSVD and FFT methods. However, FFT spectral analysis sometimes overestimate the FWHM of the (6,5) tube and show spurious modes between neighboring RBMs due to poor selectivity of neighboring modes.

The CP strength is sensitively dependent on the FWHM at the resonance energy. The FWHM of the exciton resonance of chiral tubes excited through the $E_{22}$ optical transition ranges from 50 meV to 80 meV, depending on the chirality, while those for the $E_{11}$ optical transition are as small as 30–55 meV.

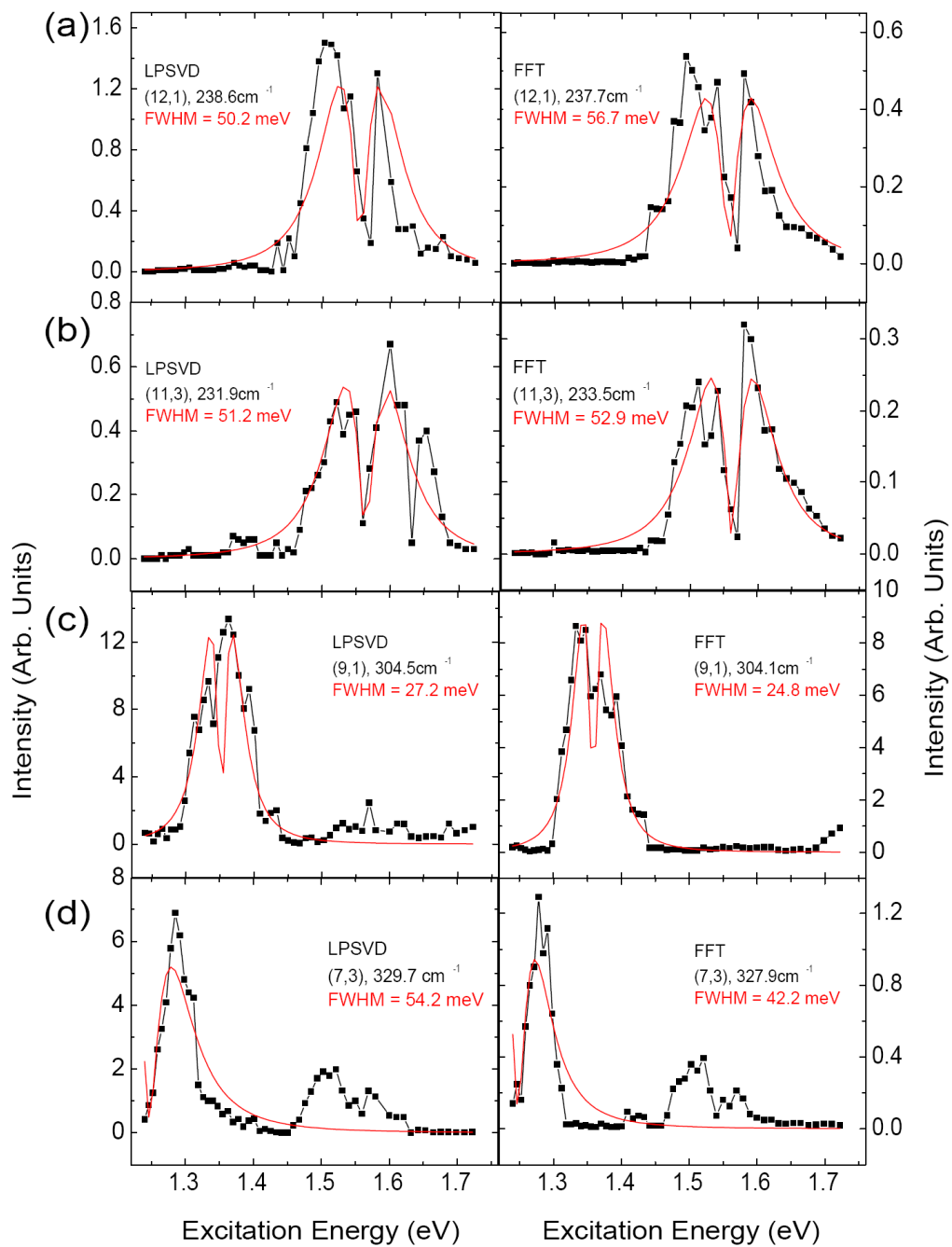

**Figure S-1** Fitting examples for calculating spectral intensity of RBMs with resonance excitation profiles for (a) (12,1), (b) (11,3), (c) (9,1), and (d) (7,3) nanotubes, obtained by LPSVD method and FFT method, respectively.